\documentclass[pre,twocolumn,showpacs,preprintnumbers,amsmath,amssymb]{revtex4}

\usepackage{color}
\usepackage{graphicx}
\begin{document}
\preprint{APS/123-QED}    
\title {Spin-1 Hopfield Model under a Random field}
\author{C. V. Morais$^1$, M. J. Lazo$^2$, F. M. Zimmer$^3$, P. R. Krebs$^1$, S. G. Magalhaes$^4$}%

\affiliation{$^1$Instituto de F\'{i}sica e Matem\'atica, Universidade Federal de Pelotas, 96010-900  Pelotas, RS, Brazil}
\email{carlosavjr@gmail.com}

\affiliation{$^2$Programa de P\'os-Gradua\c{c}\~{a}o em F\'{\i}sica - Instituto de Matem\'atica, Estat\'{\i}stica e F\'{\i}sica, 
Universidade Federal do Rio Grande, 
96.201-900, Rio Grande, RS, Brazil }

\affiliation{$^3$Departamento de Fisica, Universidade Federal de Santa Maria,
97105-900 Santa Maria, RS, Brazil}%

\affiliation{$^4$Instituto de Fisica, Universidade Federal Fluminense, 24210-346
Niter\'oi, RJ, Brazil}

\date{\today}
%

\begin{abstract}

The goal of the present work is to investigate the role of trivial disorder and nontrivial disorder in the three state Hopfield model under a Gaussian Random Field. In order to control the nontrivial disorder,  the Hebb interaction is used. It  provides a way to control the system frustration by means of the parameter $a=\frac{p}{N}$, varying from 
trivial randomness to a highly frustrated regime, in the thermodynamic limit. We performed  the thermodynamic analysis 
using the one-step replica-symmetry-breaking  mean field theory to obtain the order parameters and phase 
diagrams for several strengths of $a$, anisotropy constant and the Random Field.

\end{abstract}


\maketitle
\section{Introduction}

The role of disorder in spin systems represents a permanent source of challenging problems. For instance, random fields (RFs) and spin glass (SG) models are systems in which a richness of physical properties emerges from the disordered interactions \cite{Dotsenko,Nishimori,Young}. The interplay of these two highly nontrivial manifestations of disorder can be found in  physical systems, such as the diluted Ising-like antiferromagnets  Fe$_x$Zn$_{1-x}$F$_2$ and Fe$_x$Mg$_{1-x}$Cl$_2$ \cite{Belanger1}.
In special, for Fe$_x$Zn$_{1-x}$F$_2$, one gets a random field Ising model (RFIM)  like behavior for $x > 0.42$ and an Ising SG (ISG) for $x \sim 0.25$. For intermediate concentrations ($0.25 < x < 0.42$) one may observe both  behaviors with a crossover between them \cite{Belanger1,Belanger2}.  A more recent example of the interplay comes from disordered cerium systems, as the CeNi$_{1-x}$Cu$_{x}$ alloys (see \cite{JMMM2013} and references there in).

From the theoretical point of  view, the effects introduced by RFs are complex and nontrivial even for the most studied  RF model: the RFIM. The thermodynamic properties introduced by  RFs have remained controversial  for more than 30 years 
(for a recent review in the RFIM see e.g.  \cite{Nattermann1,Nattermann2,Dotsenko2,Belanger1}).
 For instance, several studies display broad divergences even for the RFIM  phase diagram structure \cite{Nattermann1,Nattermann2,Dotsenko2,Belanger1}. On the other hand, the effect of RFs on the SG phase has also been investigated \cite{SNA,NNCC,ANC,CN,MMN}.   In this case, random bond (RB) interaction models with an additional magnetic RF have been considered extensively. Particularly, the studies are mainly accomplished by spin-1/2 models in strong frustrated regimes (introduced by RB). 
 For example, the Sherrington-Kirkpatrick (SK) model \cite{SK} was combined with RFs following distinct types of disorder. Specifically,  the SK model with a Gaussian RF exhibits phase diagrams, in which  the SG phase decreases with the RF, at least in the mean field theory \cite{MMN}. It should be emphasized that this result was obtained within the one-step replica-symmetry-breaking (1S-RSB) approximation for the SG solution.

 However, the effects produced by RFs on phase diagrams for low
 RB frustrated regimes in spin-1 models are  almost unexplored.  Besides,
 disordered spin-1 models (like the  strong frustrated Ghatak-Sherrington \cite{ghatak} (GS) model) can exhibit a variety of interesting behaviors even without RF. For instance, their phase diagrams show a transition line that changes from a continuous phase transition to a first-order transition. Actually, the source  of this particularity
is the average number of non-magnetic states ($S=0$). Consequently, important  unexplored issues concerning the presence of RFs within a low-level of RB frustration in spin-1 models arise.  For example,  one can ask how the phase diagrams are changed by the presence of RFs within regimes of low average number of interacting states and  low-level of frustration.

Therefore, the main goal of  the present work is to analyze what kinds of phase transitions and which thermodynamic phases are present with the simultaneous  adjustment of the RF, frustration and the presence of non-interacting states. The model proposed to accomplish this task is the Spin-1 Hopfield model under a Gaussian RF. Different from the usual GS model, the spin-1 Hopfield model enables us to control the level of frustration by the Hebb interaction.  This provides a way to adjust the system disorder and frustration by means of the parameter $a=p/N$ varying from trivial disorder to a highly frustrated regime in the thermodynamic limit \cite{amit,amit2}. As a consequence, it is possible to interpolate the thermodynamics from the non frustrated $a=0$ to the strongly frustrated regime $a \rightarrow \infty$. 
In particular, these analyses  could also be a first attempt to bring helpful information in complex problems as that of cerium disordered compounds. We perform the thermodynamic analysis using the 1S-RSB mean field theory to obtain the order parameters and phase diagrams for several values of frustration level and anisotropy constant.

In addition, the present work can also shed light on the role of frustration and RF in Inverse Transitions (ITs). 
For instance, an inverse freezing (IF) occurs when increasing temperature leads the system from a liquid to a glass phase. If the increase of temperature leads to crystallization of a liquid, an Inverse Melting (IM) is obtained.
 In these kinds of phase transitions,  the phase that is usually considered the most ordered has more entropy than the disordered one. The current interest in studying  ITs is justified since it has been observed in various physical systems, as high-$T_c$ superconductors and many others \cite{0,1,2,3,4,5,6,7,8}. Moreover, the knowledge about which conditions are necessary for the occurrence of IT is a challenging issue \cite{Debenedetti}. Actually, several models have been used to  study ITs, in particular, the magnetic ones because of their simplicity \cite{9,10,11,12,13,14,15,16,17}. From those magnetic models, it is suggested that disorder from RB would be a key ingredient to produce ITs \cite{17}. On the other hand, although several works discuss the role of the RF in the IT problem \cite{16a,16b}, it is still an open issue whether or not other forms of disorder can contribute to produce ITs. Therefore, in the present work, a more detailed analysis of the role of RB and RF in ITs is performed. 

The paper is organized as follows: in Sec. II, the free energy within the 1S-RSB scheme is found.
In Sec. III, a detailed discussion of phase diagrams is presented. The last section is reserved for the
conclusions.

\section{Model}

The  Hamiltonian considered here is a Ghatak-Sherrington Hopfield model
\begin{equation}
 H= -\sum_{ij}J_{ij}S_{i}S_{j}+D \sum_{i}(S_{i})^{2}-\sum_i h_i S_i,
\label{ham}
\end{equation}
where the spin variables assume the values $S = \pm 1, 0$, the summation $(i, j)$ is over any pair of spins, $D$ is the crystal field, and $J_{ij}$ is given by 
\begin{equation}
J_{i_{}j_{}}=\frac{J}{2N}\sum_{\varrho=1}^{p}\xi_{i}^{\varrho}\xi_{j}^{\varrho}
\label{hebb}~,
\end{equation}
where $N$ is the number of spins in the system, $i,j=1,2,...,N$ are site positions, and $\xi_i^{\varrho}=\pm 1$ are independent random distributed variables following the distribution
\begin{equation}
P(\xi_{i})=\frac{1}{2}~\delta_{\xi_{i}^{\varrho},+1}+\frac{1}{2}~\delta_{\xi_{i}^{\varrho},-1}.
\label{gaussian}
\end{equation}
In addition,  the magnetic fields $\{ h_i \} $ are random variable following a independent Gaussian probability distribution:
\begin{equation}
P(h_i)=\left[\frac{1}{2\pi \Delta^{2}}\right]^{1/2} \exp\left[-\frac{1}{2\Delta^{2}}\left( h_i \right)^{2} \right].\\ 
\label{eq2}
\end{equation}

 The coupling $J_{ij}$ can be better understood by considering the local field $\phi_i=\sum_{i\neq j}J_{ij}S_j$ applied to a particular spin $S_i$ \cite{amit2}. For instance, in the spin-$\frac{1}{2}$ Hopfield model, the local field becomes $\phi_i=\xi_{i}^{1}(1+\delta_i)$ \cite{amit2} (for $J=1$), where $\delta_i$ is a random variable with variance $\langle\delta_i^2\rangle_{\xi}=\frac{p-1}{N}$. In this case, two situations can be identified in the thermodynamic limit ($N\rightarrow \infty$): $p$ finite and $p= aN$. For $p$ finite, the spins are perfectly aligned with $\xi_i^1$, which gives a free energy as that one of an usual ferromagnet \cite{binder,fradkin}. For $p= aN$, the term $\delta_i$ becomes important and the alignment can be destroyed. Particularly, $a\rightarrow \infty$ corresponds to a strongly frustrated regime given by a random Gaussian distributed $J_{ij}$ \cite{provost}, as in the SK model \cite{SK}. As discussed before, the spin-1 Hopfield model represents a more complex case, in which besides $a$, the non-interacting spin states can also be favored.

The free energy is obtained by using the replica method
\begin{equation}
\beta  f =
-\lim_{N\rightarrow\infty}\lim_{n\rightarrow 0} 1/(nN)(\left[\langle\langle {{Z^n}}\rangle\rangle_{\xi}\right]_h-1)
\label{replica}
\end{equation}
where $\langle\langle ... \rangle\rangle_{\xi}$ means the configurational averaged over $\xi$ and $\left[...\right]$ denotes the configurational average over the random fields. We follow closely the procedure used in Ref.~\cite{Nobre1} to calculate the average over the random field, which means that the average over the $h_i$. It can be obtained as,
\begin{equation}
\begin{split}
\left[\langle \langle Z^n \rangle \rangle_{\xi}\right]_h
=\int 
\prod_i dh_{i}P(h_i) \langle \langle Z^n \rangle \rangle_{\xi}. \\
\end{split}
\label{eq3}
\end{equation}
After performing the 
average $\left[  . . \right]_h$, the partition function becomes
\begin{equation}
\left[\langle \langle Z^n \rangle \rangle_{\xi}\right]_h 	=
\langle \langle \mbox{Tr}_{s_{\alpha}}
\exp[A_{f}^{\alpha} + A_{SG}^{\alpha}] \rangle \rangle_{\xi}
\label{zn1}
\end{equation}
with

\begin{equation}
 A_{f}^{\alpha}= -\beta D\sum_{i}\sum_{\alpha}(S_{i}^{\alpha})^{2} +  \frac{(\beta\Delta)^2}{2}\sum_i
( \sum_{\alpha} S_{i}^{\alpha} )^2 
\end{equation}
\begin{equation}
A_{SG}^{\alpha}= \frac{\beta J}{2N}\sum_{\varrho=1}^{p}\sum_{\alpha=1}^{n}(\sum_{i}
\xi_{i}^{\varrho}S_{i}^{\alpha})^{2} - \frac{\beta Jp}{2N}\sum_{i}\sum_{\alpha=1}^{n}(S_{i}^{\alpha})^{2}
\label{asg},
\end{equation}
with $\alpha$ denoting the replica index.

The average  $\langle\langle ... \rangle\rangle_{\xi}$ over $Z^n$ {given} in Eq. (\ref{zn1}) is discussed in detail in Appendix A. In the present work, the  one-step replica-symmetry-breaking (1S-RSB) ansatz is adopted (see Appendix B), which results in the following equation for the free energy 

\begin{equation} 
\begin{split}
& \beta f= B_a R_0
 +\frac{\beta J m^2}{2} 
-\frac{1}{2} \frac{\beta J a q_0}{Q_0}
+\frac{a}{2}\ln [Q_1]
\\&+\frac{a}{2x}\ln \frac{Q_0}{Q_1}
 -\frac{1}{x}\int Dz\langle\langle\ln \int Dv\left(K(z,v|\xi)\right)^x\rangle\rangle_{\xi}
\label{eq1}
\end{split}
\end{equation} 
where $B_a= \frac{\beta^2 J^2 a}{2}$, $R_0=\bar{r}\bar{q}-(1-x)r_1q_1-xr_0q_0$, $Q_0=1-\beta J[\bar{q}-q_1+ x(q_1-q_0)]$, $Q_1= 1-\beta J(\bar{q}-q_1)$, with
\begin{equation}
K(z,v|\xi)=1 + 2 \mbox{e}^{\gamma}\cosh\bar{H}(z,v,\xi),
\label{Kzv}
\end{equation}
\begin{equation}
\gamma = \frac{\beta J a}{2} [\beta J (\bar{r}-r_1)-1 ] - \beta D,
\end{equation}
and
\begin{equation}
\begin{split}
\bar{H}(z,v|\xi)&=\beta J[\sqrt{ar_0 + (\Delta/J)^2}z+\sqrt{a(r_1-r_0)}v+\xi m].
\end{split}
\label{eq20}
\end{equation}
 The order parameters $m$, $q_{0}$, $q_{1}$ and $\bar{q}$ are defined in Appendix B. In the present treatment, $m$ represents  the ``magnetization'' that characterizes the Mattis state ($m=\langle\langle\xi <S>\rangle\rangle_{\xi}$ with the thermodynamic average $<\cdots>$) \cite{amit}. $\delta\equiv q_1-q_0$ is the Parisi 1S-RSB order parameter \cite{Parisi}. $\bar{q}$ is related to the average  number $n_0=1-\bar{q}$ of nonmagnetic states in the sites, which also reflects the capability of the	 sites to interact or not \cite{Castillo}.

In particular, the elements of matrix r are given by
\begin{equation}
 r_0=\frac{q_0}{Q_0^2},~ r_1-r_0=\frac{q_1-q_0}{Q_1 Q_0},~ \bar{r}-r_1=\frac{1}{\beta J Q_1}.
\label{r01}
\end{equation}
The average over $\xi$ in the free energy can be done using the parity properties of the functions dependent on $z$ and $v$. The entropy $s= -\partial  f/ \partial T$ can be derived from the free energy.

\section{Results and Discussion}

\begin{figure}[htb]
\begin{center}
\includegraphics[angle=270,width=8.5cm]{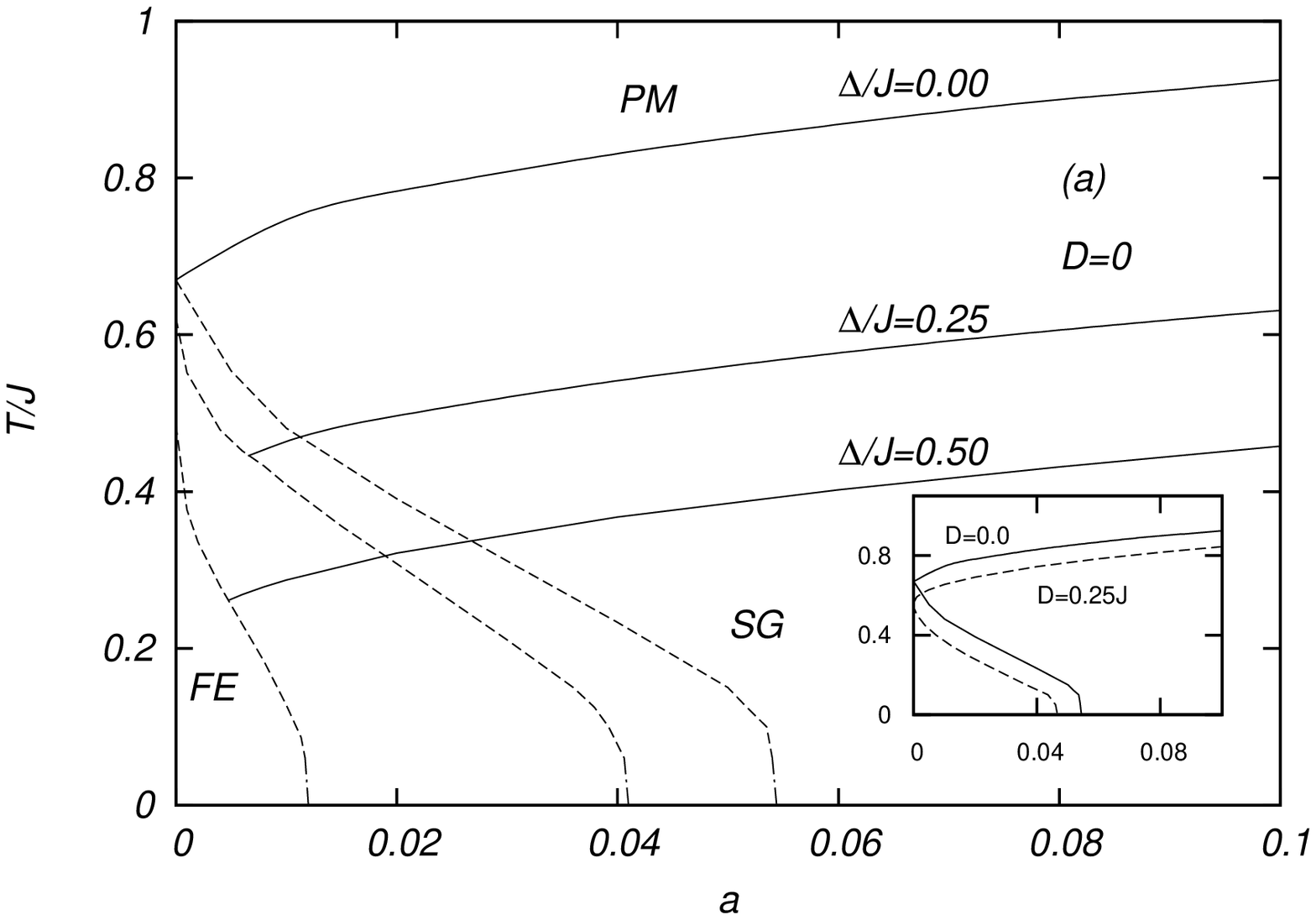}
\includegraphics[angle=270,width=8.5cm]{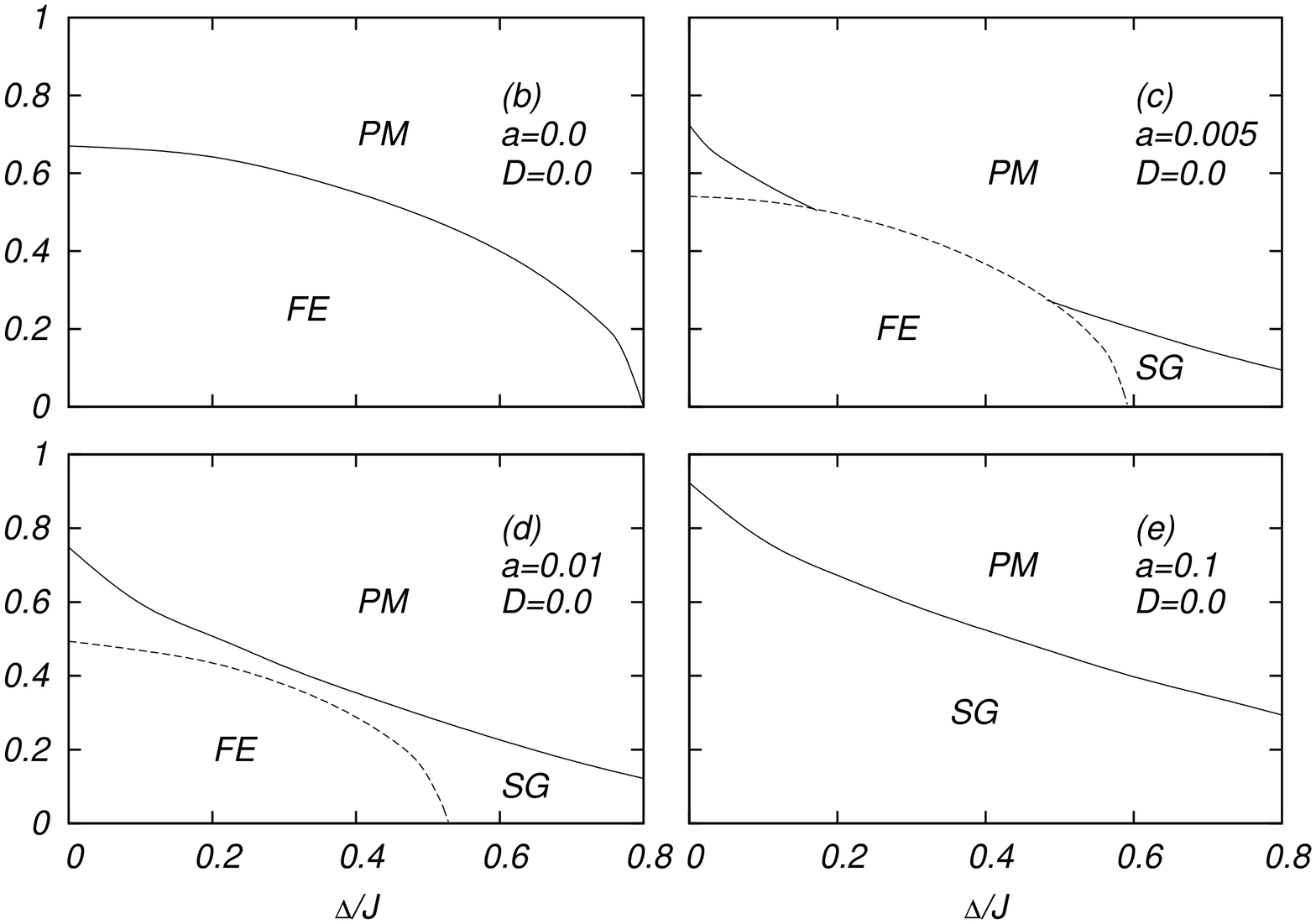}
\end{center}
\caption{Fig. (a) exhibits phase diagrams $T/J$ {\it vs} $a$ for $D/J=0.0$ and several values of $\Delta/J$. Figs. (b)-(e) show phase diagrams $T/J$ {\it vs} $\Delta/J$ for several values of $a=p/N$. Full and dashed lines represent
 the continuous and the first order phase transitions, respectively.}
\label{fig1}
\end{figure}

This section displays phase diagrams $T/J$ versus $a$ and $T/J$ versus $D/J$ that are built from numerical
solutions of the order parameters for several values of $\Delta/J$. For the numerical results, $J=1$ is used. 
In these, the SG phase is characterized by RSB solution ($\delta\equiv q_1-q_0>0$) with zero magnetization
($m=0$). The Mattis state (FE) occurs when the RS solution ($q=q_1=q_0$) is found with
$q>0$ and $m>0$.  The paramagnetic (PM) phase is characterized by $m=0$ with RS solution.
Particularly the RF induces the correlation given by $q$ in the PM phase, but without  breaking  the
replica symmetry ($\delta=0$, $q>0$ with $m=0$). The behavior of the entropy $s$ and  the average number of non-magnetic sites $n_0$ 
\cite{Castillo} are also discussed in this section.

\begin{figure*}[htb] 
\begin{center}
\includegraphics[angle=270,width=16cm]{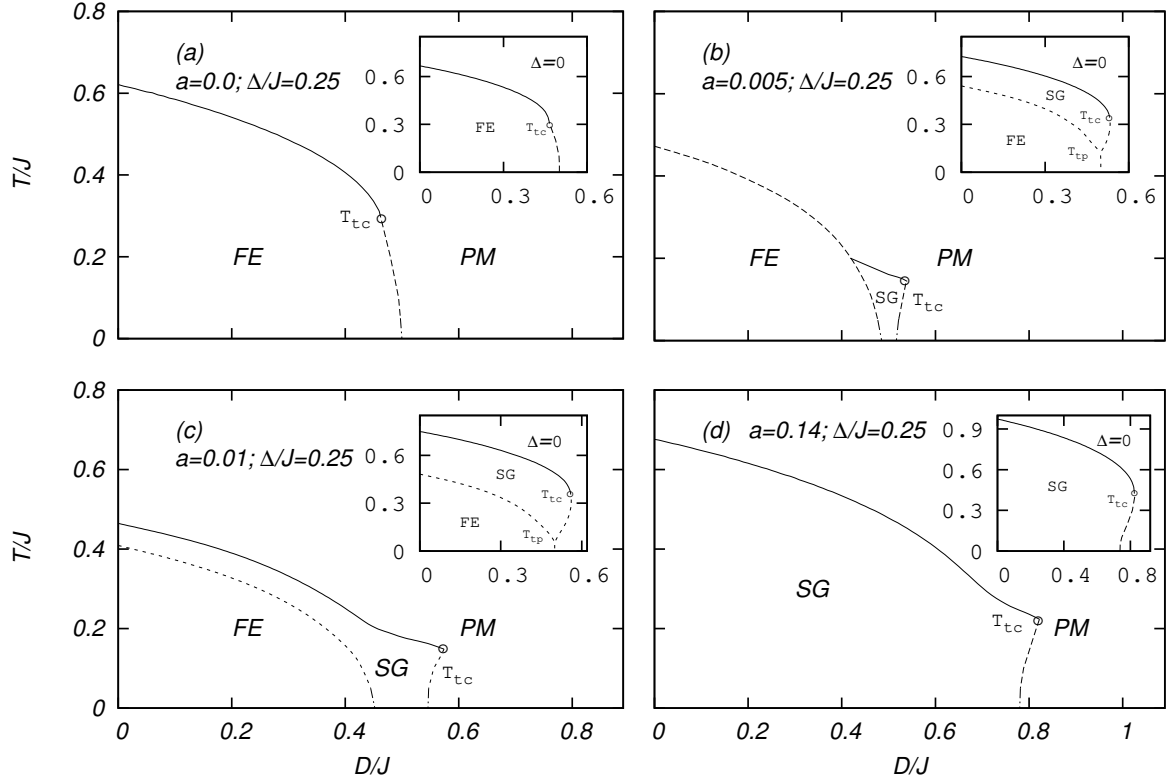}
\end{center}
\caption{Phase diagrams $T/J$ {\it vs} $D/J$ for $\Delta/J=0.25$ and  several values of $a=p/N$. Full and dashed lines represent 
 the continuous and the first order phase transitions, respectively.  $T_{tc}$ and $T_{tp}$ represent the tricritical and triple points, respectively. 
}
\label{fig2}
\end{figure*}

\begin{figure*}[htb]
\begin{center}
 \includegraphics[angle=270,width=8.6cm]{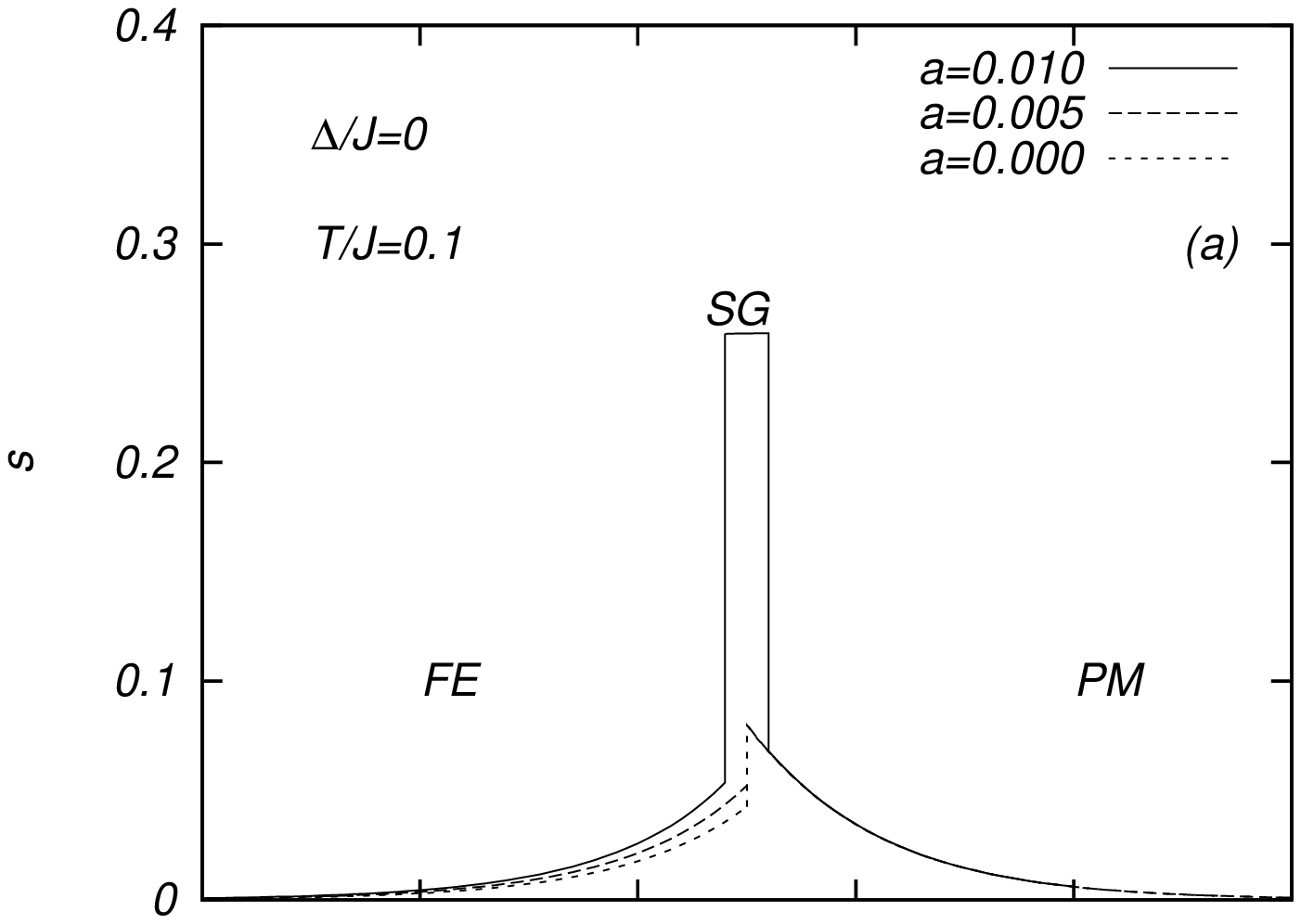}
 \includegraphics[angle=270,width=8.6cm]{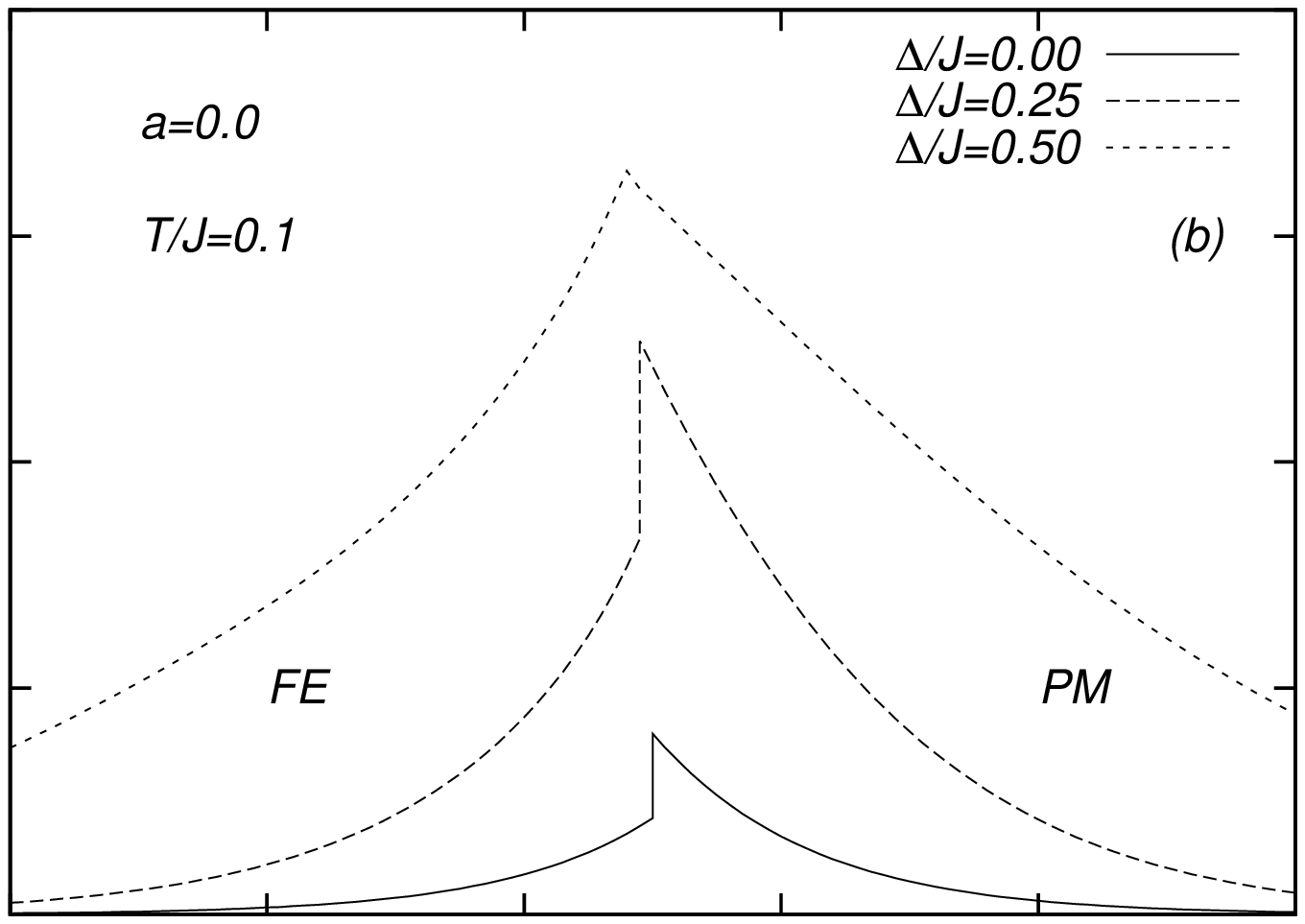}
 \includegraphics[angle=270,width=8.6cm]{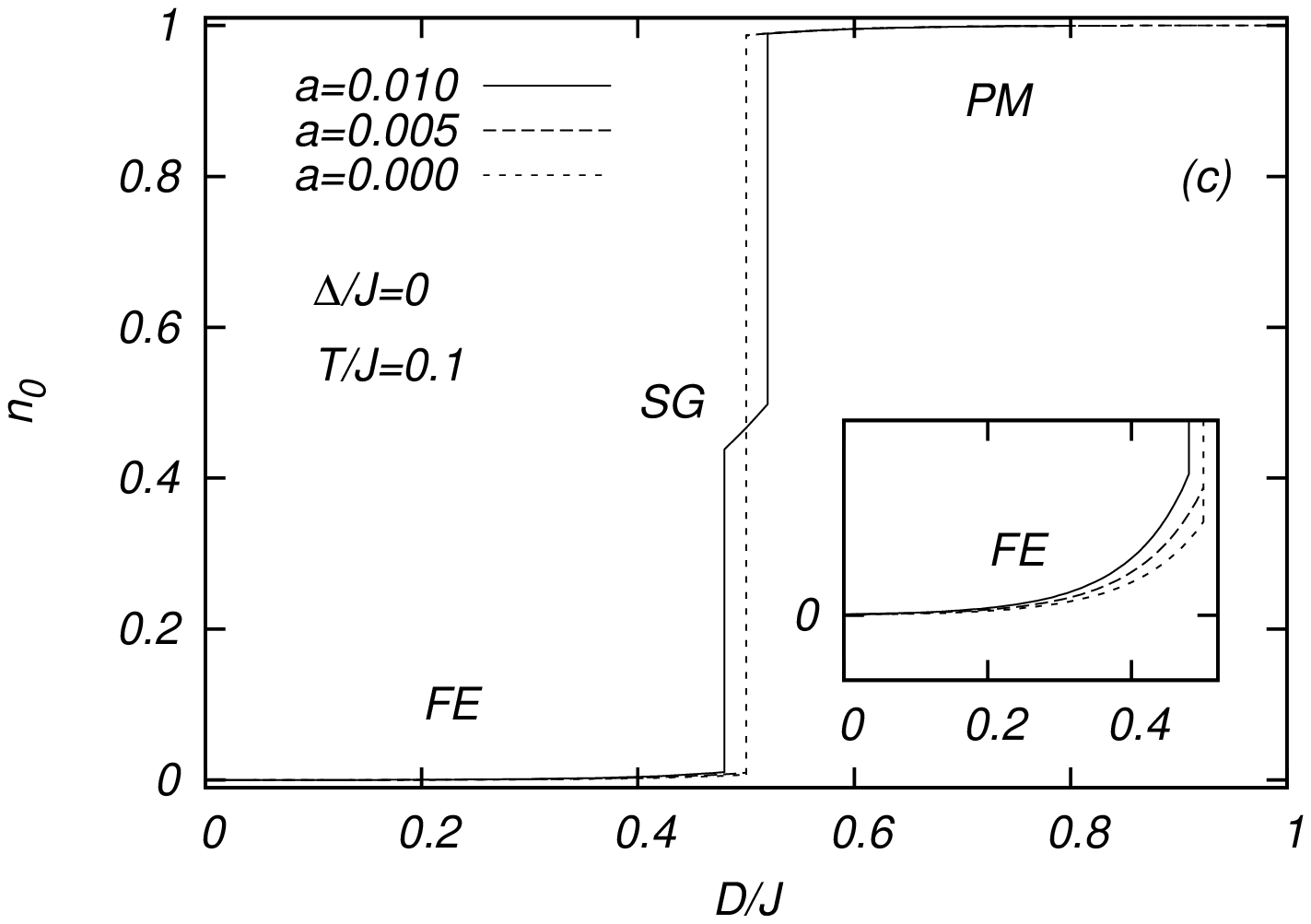}
 \includegraphics[angle=270,width=8.6cm]{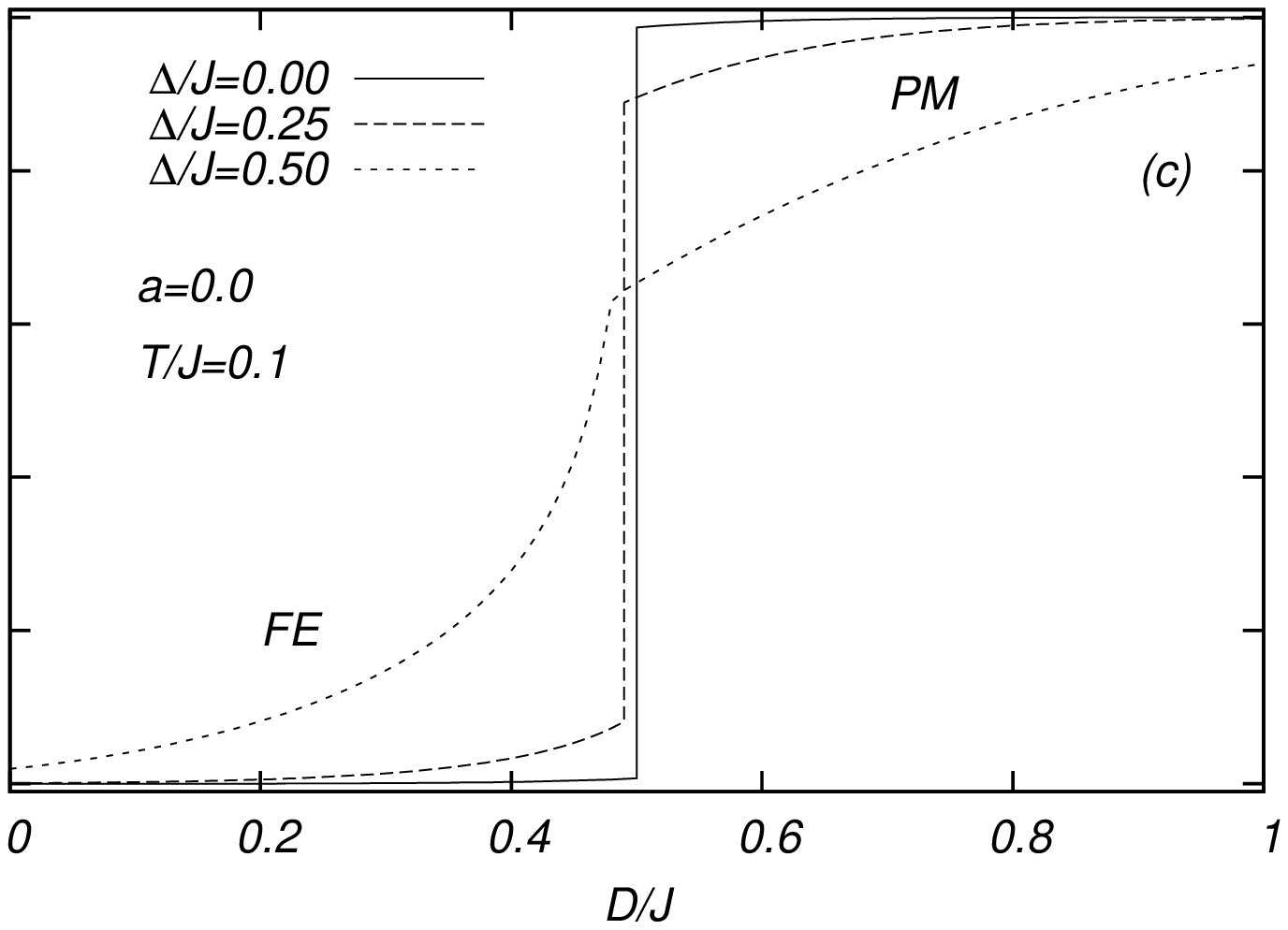}
 \end{center}
\label{fig3}
 \caption{Fig. (a) shows the Entropy vs $D/J$ for $\Delta=0$ and several values of $a=p/N$. Fig. (b) exhibits the entropy vs $D/J$ for $a=0$ and several values of $\Delta/J$. The  behavior of the average number occupation of non-magnetics sites $n_0$ vs $D/J$ is shown in Figs. (c) and (d). }
\end{figure*}

\begin{figure*}[htb]
\begin{center}
 \includegraphics[angle=270,width=8.6cm]{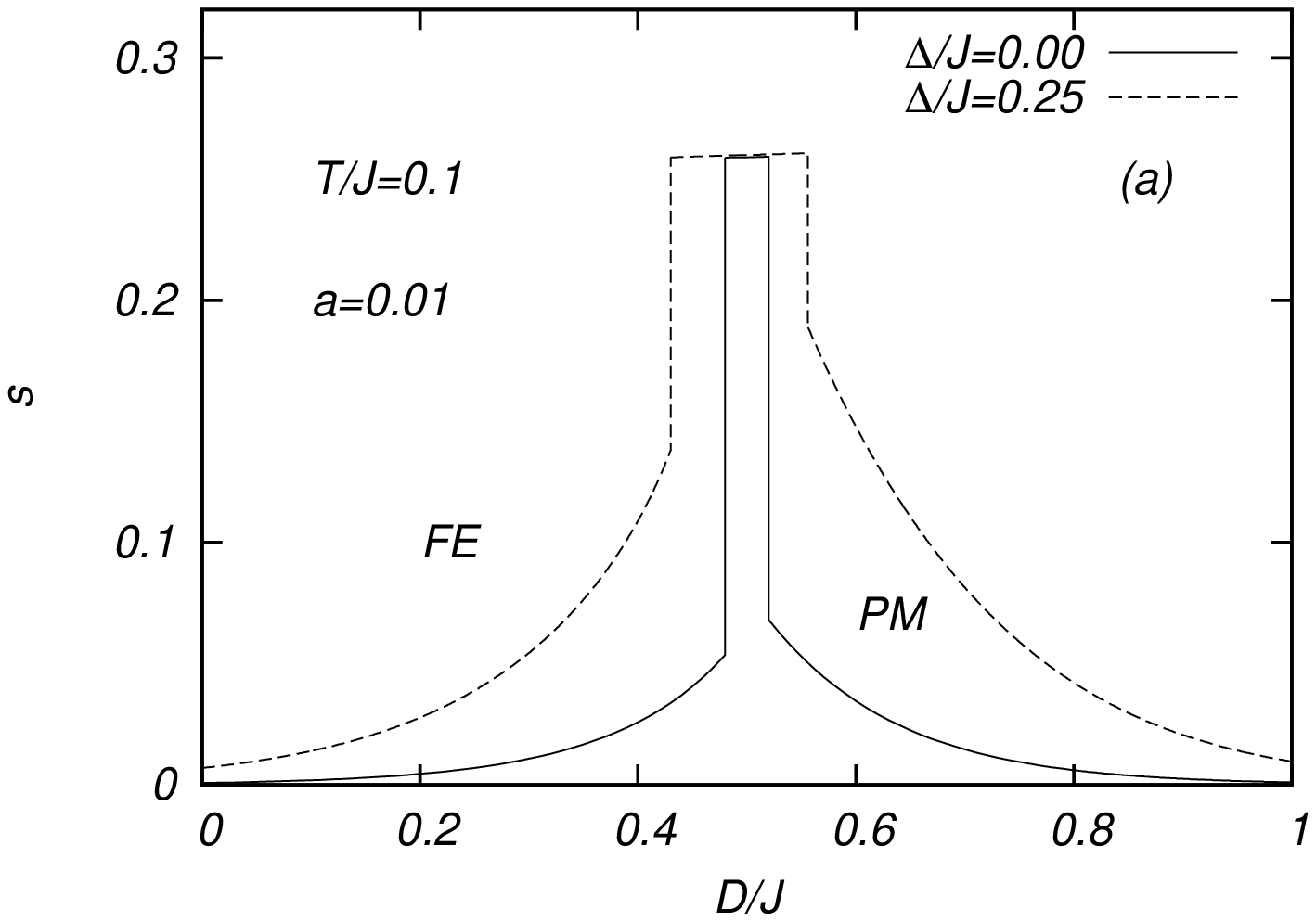}
 \includegraphics[angle=270,width=8.6cm]{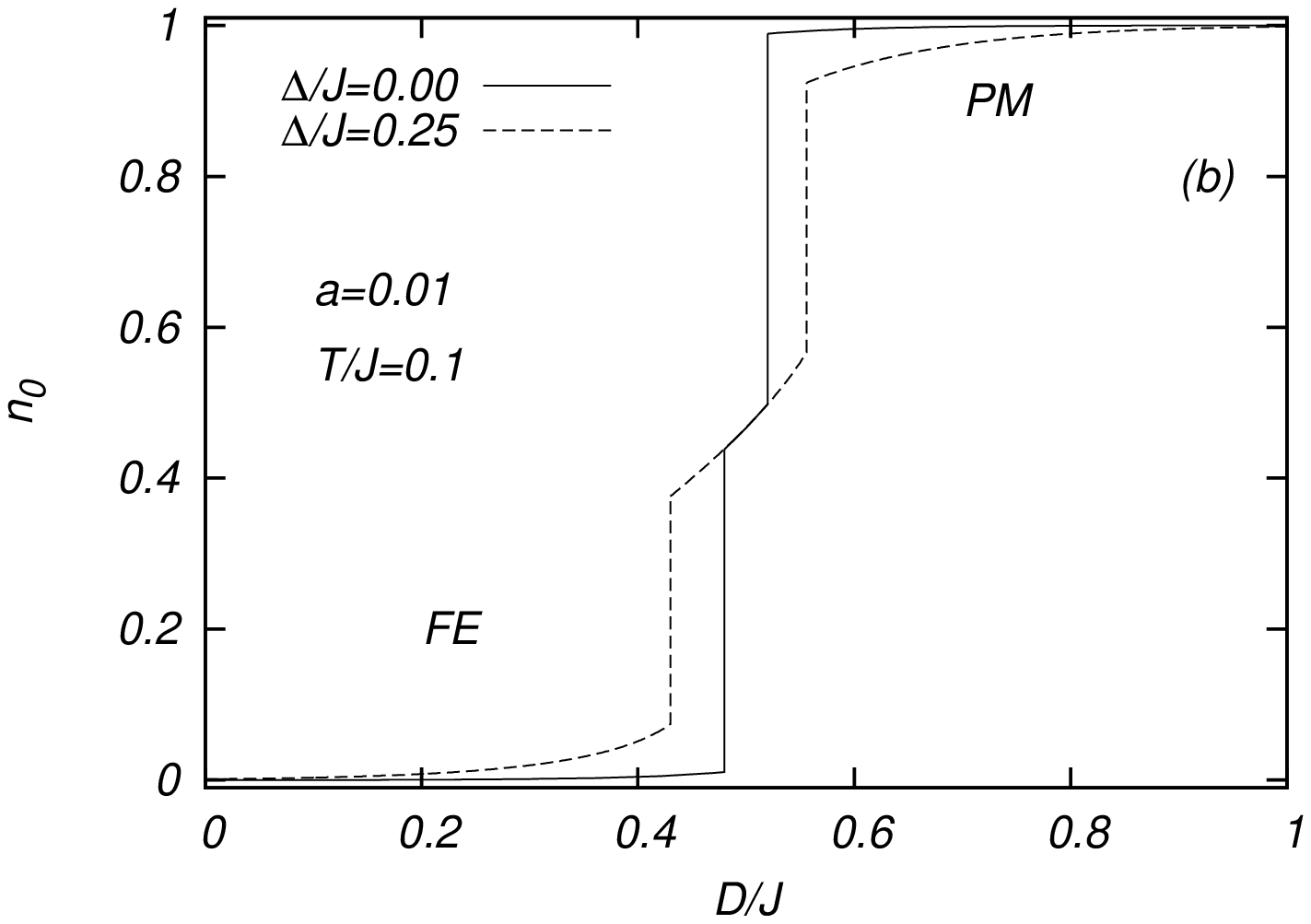}
 \end{center}
\label{fig4}
 \caption{Fig. (a) shows the Entropy vs $D/J$ for $\Delta=0,0.25$ and  $a=0.01$.  The bevahior of the average number occupation of non-magnetics sites $n_0$ vs $D/J$ is shown in Fig. (b). }
\end{figure*}

Effects of trivial and nontrivial disorders on phase boundaries are  analyzed in detail. Here, these disorders are controlled by tuning $a=p/N$ and $\Delta$, which are associated with the degree of frustration and the variance of Gaussian RF (see Eq. \ref{eq2}), respectively. However, before going into details of phase diagrams with combined effects of disorder, a summary of the results for  special situations are 
presented.  For instance, it is known that the spin-1 Hopfield model in absence of RFs ($\Delta=0$) exhibits a RS phase diagram in which the FE phase, as well both the  FE/SG phase transition and the freezing temperature $T_f$ decrease as the fraction of nonmagnetic sites $D$ increases \cite{bolle}. This behavior is also observed in our 1S-RSB results presented in the inset of  \ref{fig1}(a).  
On other hand, when $D\rightarrow -\infty$ the behavior of the spin-1/2 Hopfield model is recovered. In this case, the increase of $\Delta$ can reduce both the SG and FE phases, where FE/SG transition disappears for  a critical value of $\Delta$ \cite{qiang}.   
Another important limit  is $a\rightarrow \infty$, in which the results for the high frustrated SK $J_{ij}$-type interactions can be obtained \cite{provost}. In particular, one can reproduce the  GS phase diagrams  (see references \cite{16a,16b}).

However, the simultaneous presence of both disorders in low frustrated regimes of the spin-1 Hopfield model  produces even more interesting results. For instance, Fig. \ref{fig1}(a) exhibits the case with $\Delta=0$, where the SG/PM  critical frontier $T_f$ increases with the parameter $a$. For small values of $a$, a first order SG/FE phase transition line $T_{1c}$ is also found as the temperature diminishes.  Particularly, the $T_{1c}$ corresponds to the limit of the global stable FE solution. Moreover,  
this result is qualitatively the same as that found in Ref. \onlinecite{amit2}.  Nevertheless, the presence of the RF produces  important changes in the phase boundaries. For example,  $T_f$ and $T_{1c}$ decrease as $\Delta$ increases.  Mostly important, a direct first order line transition  appears between the PM and FE phases for $\Delta/J=0.25$ and $0.5$. It indicates that the RF affects the FE and SG  phase  boundaries differently as can be seen in Figs. \ref{fig1} (b)-(e). The $a=0$ case shows for small  $\Delta/J$ a continuous (second order) phase transition (called  $T_{2c}$) between FE and PM phases.  Increasing $\Delta/J$ to a critical value,  the FE phase is completely suppressed. The increase of $a$ produces a more complex scenario  with arising of the SG phase.  Furthermore,  
 different from the FE phase, the SG one is not completely suppressed  by the RF.
For instance, the $T_f$  decreases monotonically as $\Delta/J$ increases.  In addition, the nature of the FE phase boundary is changed from  continuous to first order (see Fig. (c)).  Moreover, for further increase of $a$,
Figs. (d)-(e) indicate  the complete suppression of the FE. To summarize  the results found in Fig. 1, the $a$ and RF 
can completely suppress the FE state. On the contrary,  it is observed that the parameter $a$ favors the SG phase, while the  RFs undermine it. 

The effects on the phase boundaries coming from  tuning $D$  are analyzed in more detail 
in Fig. \ref{fig2}. For instance, Fig. \ref{fig2}(a) exhibits the case where only trivial disorder is present ($a=0.0$, $\Delta=0.25J$). For low $D/J$, there is only a continuous FE/PM phase transition, with no SG phase (see Fig. \ref{fig2}(a) and  its
inset for comparison). Increasing $D$,  a tricritical point $T_{tc}$ appears and the FE/PM phase transition becomes first order.
Fig. \ref{fig2}(b) displays a phase diagram where nontrivial disorder is  also present ($a=0.005$  and $\Delta=0.25J$). In this case, the intensity of $\Delta$ is high enough to prevent the occurrence of SG phase at $D=0$ (see inset of Fig. \ref{fig2}(b) for $\Delta=0$). However, the  first order  PM/FE phase transition is replaced by a continuous PM/SG transition when $D$ increases, where the SG phase appears for $0.42J\leq D \leq 0.56 J$.  From Figs. \ref{fig2}(c)-(d), it is observed a smooth decreasing of $T_f$ which can appear  at larger values of $D/J$ when the RF is present (see insets for comparison). Another point that should be emphasized is the suppression of the triple point  $T_{tp}$ which appears only for low frustration levels. It does not appear for the $a=0$ case or when  $a$ is high enough. The additional disorder coming from the RF tends to suppress the $T_{tp}$.

It is also important to remark that the first order  boundaries from previous results are located  by comparing the free energies, where the ``stable'' phase is that one with lower free energy. 
 In this case, it should be emphasized that the location of the first-order transitions is not very dependent on the RSB scheme \cite{16a,16b,14}.  In particular, the increase of $D$ results in a reentrant first order SG/PM  phase boundary whenever frustration is present. This reentrant behavior can be an indication of inverse freezing (IF). On the other hand, no reentrant behavior is found in the FE/PM phase boundaries,  which means absence of inverse melting (IM) \cite{9}. To clarify these points, a detailed analysis of the entropy and behavior of the average number of non-magnetic states $n_0$ is displayed in Fig. 3. Fig. 3(a) shows the behavior of entropy $s$ as a function of $D/J$ for $T/J=0.1$ and several values of $a$ and $\Delta=0$.  Particularly, this result shows the effects of different disorders 
 on the entropy of the FE, SG and PM phases. For $a=0$, the entropy of the FE phase is smaller than the PM one at the first order transition.   
There is  a small increase in the entropy of the FE phase for higher values of  $a$.
However,  the entropy of the FE phase  can still be found below the PM one at the transition. 
On the other hand, the entropy of the PM phase remains almost unaltered with increasing of $a$.  Mostly important, these results indicate that the parameter $a$ favors  the FE phase entropically. Nevertheless, an  IM does not  occur because a highly entropic SG phase appears with the increasing of  $a$,  replacing the FE phase.  This SG phase appears as more entropic than the PM one, which is the conditions for  IT (an IF in this case). Finally, Fig. 3(b) shows  the  entropy behavior for $a=0$ as $\Delta/J$ increases. This result indicates that the RF increases the entropy of both FE and PM phases. As a consequence, no IM can be found 
 too.

The behavior of the average number of non magnetic sites $n_0$ as a function of $D/J$ for $T/J=0.1$ and several values of $a=p/N$ and $\Delta/J$ are shown in Figs 3(c) and 3(d). The limits $n_0=0$ and $n_0=1$ can be related to the interacting  and the no interacting regimes, respectively. 
The effects of  $a$ on $n_0$  in the FE, SG and PM phases are shown in Fig. 3(c). For instance,  $n_0$  
 presents a small increase with $a$ in the FE phase.
In addition, $n_0$ increases to an intermediate value when the SG phase appears for $a=0.01$.  On the contrary, $n_0$ in the PM phase for larger $D/J$ remains unaltered. As a consequence the entropy of this PM phase remains low even for higher values of $a$. In Fig. 3(c) the disorder related to the RF is increased.  It should be remarked that
the effects of $\Delta/J$  on $n_0$  differ from those obtained by increasing $a$. In particular,  $n_0$ in  low temperature PM phase presents
a  clear decreasing  as $\Delta/J$ increases.  This result shows that 
the RF disfavors non-interacting states in this PM phase.  
In that sense, the RF increases the entropy of the PM phase  at a lower temperature. Therefore, it tends to suppress 
the reentrance in the SG/PM phase boundary. 

The  effect  of tuning $a$ and $\Delta$ on  the  entropy and the $n_0$  are analyzed in Fig. 4. 
In Fig. 4(a) the entropy of the FE, SG and PM phases increase with the RF. However, the increase 
of the  SG entropy  is less pronounced than those observed in the FE and PM phases. Particularly, Fig. 4(b) indicates that the $n_0$ remains at intermediate values even for larger values of $D/J$. It occurs because the RF couples with the interacting spin states.  As a consequence,  the SG phase  appears for larger $D/J$. In addition, $n_0$ of the SG phase does not present a significant increasing with the RF. Such behavior of the $n_0$ could be the reason for the SG entropy to present only 
a very small increase with $\Delta/J$.

\section{Conclusion}

The present work studied the  Spin-1 Hopfield Model under a Gaussian random field (RF). This study was carried out within the 1S-RSB scheme using a mean field approximation. The proposed approach allows to explore the relation between frustration (related to the RB disorder) and  random field in the presence of noninteracting states (coming from 
the crystal field $D$). As a consequence,  new  information concerning the physical  
properties related to the interplay of the mentioned effects can be obtained.

In our model, the degree of frustration and the RF disorder are controlled by tuning $a=p/N$ and $\Delta$, where the last one is associated to the variance of Gaussian RF. For very low values of $\Delta$ the RB dominates and one can find FE or SG phase depending on the frustration level. On the other hand, for very high values of $\Delta$ the RF dominates and both FE and SG phases are suppressed within the 1S-RSB scheme. 
The increasing of $a$ produces a complex scenario with arising of the SG phase and FE phase suppression. In fact, while the FE phase disappears quickly by introducing disorder coming from both RB and RF, the SG is reinforced by increasing the frustration level and  it is monotonically suppressed by increasing  $\Delta$. 

 A  raise in the number of noninteracting states   $S=0$ by tuning the crystal field $D$ increases the complexity of the thermodynamics obtained from the proposed model. For instance, the  FE phase is intensively affected by the crystal field, since the occupation of interacting states is fundamental in defining the orientation of spins. Particularly, the FE phase is suppressed  by increasing $D/J$. 
Moreover, in the presence of frustration, the FE phase can be replaced by a SG phase  even without RF. However, the changes produced by the interplay of the RB, RF and $D$ are far from trivial. Deep changes can occur in the structure of the phase diagrams depending on the combination of the mentioned parameters. The most notable consequence  occurs at sufficiently low frustration levels. In such cases, no SG phase can be found in the presence of RF when there are a few number of $S=0$ states. Due to the competing effects between the RF (coupling with interacting states) and $D$ (favoring $S=0$ states) the SG phase can appear at large D. Although this SG phase appears,
 higher $D/J$ values still suppress the SG phase by  increasing the number of $S=0$ states.

Finally, another goal is to respond whether a complex form of disorder, the nontrivial one defined by the presence of frustration,  is an essential ingredient to produce Inverse Transitions (ITs). For finite $a$, we can superpose  this disorder with effects coming from the Gaussian RF. This procedure allows to probe the role of both RF and frustration on 
ITs. From our results it is shown that only nontrivial disorder can cause ITs even for very  low level of frustration. Moreover,  at lower temperature, our results clearly display that the main effect of nontrivial disorder is to increase the entropy of the ordered phases, particularly, of the SG one.

We also hope that the present results showing the interplay between SG and RF could be helpful for theoretical description of disordered magnetic systems as, for instance, 
the $CeNi_{1-x}Ni_{x}$ alloys \cite{Coqblin1,Coqblin2}.

\section*{Acknowledgments}

This work has been partly supported by CNPq, CAPES, FAPERJ and FAPERGS (Brazilian agencies).

\appendix

\section{The average over $Z^n$ }\label{apendicea}

In this appendix,  the averaging procedure of the partition function is introduced following closely 
Ref. \onlinecite{amit}.  The first term in the action $A^{\alpha}_{SG}$ (see Eq. (\ref{asg})) can be 
linearized by a Hubbard-Stratonovich transformation by introducing $n \times p$ auxiliary fields 
$m_{\varrho}^{\alpha}$ which are splitted in two subsets with $n \times (p-l)$ and $n \times l$ terms. 
Therefore,
 \begin{equation}
\begin{split}
&\exp(A_f^{\alpha}+A_{SG}^{\alpha})=\exp\left[ A_D^{\alpha} + A_{RF}^{\alpha}
\right]
\\ &\times 
\int_{-\infty}^{\infty} D m_{\nu}^{\alpha}  
\exp\left\lbrace \tau \sum_{\nu=1}^{l}\sum_{\alpha}\left[ -\frac{1}{2}(m_{\nu}^{\alpha})^2 + \eta_{i,\alpha}^{\nu}
\right] \right\rbrace \\ &\times
\int_{-\infty}^{\infty} D m_{\varrho}^{\alpha}  \exp\left\lbrace \tau \sum_{\varrho=l+1}^{p}\sum_{\alpha}\left[ -\frac{1}{2}(m_{\varrho}^{\alpha})^2  + \eta_{i,\alpha}^{\varrho}
\right] \right\rbrace,
\label{mu}
\end{split}
\end{equation}
where $A_D^{\alpha}=(-\frac{\beta Jp}{2N}-\beta D)\sum_{\alpha} \sum_{i}(S_{i}^{\alpha})^{2}$, $A_{RF}^{\alpha}= \frac{(\beta \Delta)^2}{2}\sum_{i} (\sum_{\alpha} S_{i}^{\alpha})^{2}$,  $\tau=\beta J N$,  $\eta_{i,\alpha}^{\nu(\varrho)}=\frac{1}{N}\sum_{i}\xi_{i}^{\nu(\varrho)}S_{i}^{\alpha}m_{\nu(\varrho)}^{\alpha}$  and $D m_{\nu(\varrho)}^{\alpha}=\prod_{\nu(\varrho)}\prod_{\alpha}\frac{dm_{\nu(\varrho)}^{\alpha}}{\sqrt{2\pi}}$.

It is assumed that the relevant contributions come from $m_{\nu}^{\alpha}$ which are order unity, while $m_{\varrho}^{\alpha}$ is of order $1/\sqrt{N}$. Therefore, the  average over the $p - l$ independent random variables $\xi_{i}^{\varrho}$ can be done using $P(\xi_{i}^{\varrho})$ given in Eq. (\ref{gaussian}), which results in:
\begin{equation}
\langle\langle \exp\left[ M_{u}^{\alpha} \right] \rangle\rangle_{\xi}= \nonumber \\ 
\exp%
\sum_{i}\sum_{\varrho=s}^{p} \ln\left( \cosh (\beta J \sum_{\alpha}S_{i}^{\alpha}m_{\varrho}^{\alpha})\right) .
\label{mu1}
\end{equation}
with $M_{u}^{\alpha} = \beta J\sum_{\varrho=l+1}^{p}\sum_{\alpha}(\sum_{i}\xi_{i}^{\varrho}S_{i}^{\alpha})m_{\varrho}^{\alpha}$. The argument of the exponential in the right hand side of Eq. (\ref{mu1}) can be expanded up to second order in $m_{\varrho}^{\alpha}$. The result is a quadratic term of the spins variables $S_{i}^{\alpha}$ in the last exponential of Eq. (\ref{mu}). This term can be linearized by introducing the spin glass order parameter $q_{\alpha\beta}$ using the integral representation of the delta function as
$
\int_{-\infty}^{\infty}\frac{dr_{\alpha\beta}^{'}}{2\pi} \exp\left[ i r_{\alpha\beta}^{'}(q_{\alpha\beta}-\frac{1}{N}\sum_{i}S_{i}^{\alpha}S_{i}^{\beta})\right]
= \delta(q_{\alpha\beta}-\frac{1}{N}\sum_{i}S_{i}^{\alpha}S_{i}^{\beta}). 
\label{mu2}
$
Therefore, the exponential involving $m_{\varrho}^{\alpha}$ in Eq. (\ref{mu}) can be written as: 
\begin{equation}
\begin{split}
&\exp\left\{\beta N
\sum_{\varrho=l+1}^{p}\sum_{\alpha}\left[-\frac{1}{2}(m_{\varrho}^{\alpha})^2
+\frac{1}{N}(\sum_{i}\xi_{i}^{\varrho}S_{i}^{\alpha})m_{\varrho}^{\alpha}\right]\right\}\\&=
\int_{-\infty}^{\infty} 
\prod_{\alpha\beta} \frac{dq_{\alpha\beta}
d\tilde{r}_{\alpha\beta}}{2\pi} 
\exp \left[  \frac{\beta}{2} \sum_{\varrho=l+1}^{p} m_{\varrho}^{\alpha}\Lambda_{\alpha\beta} 
m_{\varrho}^{\beta} G_{\alpha\beta} \right] ,
\label{e20}
\end{split}
\end{equation}
with
$G_{\alpha\beta}=i\sum_{\alpha\beta}{\tilde{r}}_{\alpha\beta}(q_{\alpha\beta}-\frac{1}{N}\sum_{i}S_{i}^{\alpha}
S_{i}^{\beta})$
where the matrix element
$\Lambda_{\alpha\beta}=(1- \beta q_{\alpha\alpha})\delta_{\alpha\beta} + 
\beta q_{\alpha\beta} (1- \delta_{\alpha\beta})$. 

Inserting Eqs. (\ref{e20}) 
into Eq. (\ref{mu}), the
$m_{\varrho}^{\alpha}$ fields can be integrated to give:
\begin{equation}
\begin{split}
&\langle\langle \exp (A_{f}^{\alpha} + A_{SG}^{\alpha} ) \rangle\rangle_{\xi} =
\exp\left[ A_D^{\alpha} + A_{RF}^{\alpha}\right]
\\ &\times 
\langle\langle \int^{+\infty}_{-\infty} Dm_{\nu}^{\alpha} 
\exp\left\{\tau \sum_{\nu=1}^{l}\sum_{\alpha} 
\left[ -\frac{1}{2}(m_{\nu}^{\alpha})^{2}+
\eta_{i,\alpha}^{\nu}
\right]\right\}
\rangle\rangle_{\xi} 
 \\ &\times
\int^{\infty}_{-\infty}\prod_{\alpha\beta}
\frac{dq_{\alpha\beta}{\tilde{r}}_{\alpha\beta}}{2\pi}
\exp\{ G_{\alpha\beta}-
\frac{1}{2}(p-l)Tr\ln\underline{\underline{\Lambda}}) \}.
\label{e22}
\end{split}
\end{equation}

Assuming $l=1$ in Eq. (\ref{e22}), the averaged partition function is given as

\begin{equation}
\begin{split}
&\langle\langle Z^n \rangle\rangle_{\xi}=
\int^{\infty}_{-\infty}Dm^{\alpha}_{1}\int^{\infty}_{-\infty}\prod_{\alpha\neq\beta}
\frac{dq_{\alpha\beta}d{\tilde{r}}_{\alpha\beta}}{2\pi}
\prod_{\alpha}\frac{dq_{\alpha\alpha} d\tilde{r}_{\alpha\alpha}}{2\pi} 
\\ &\times \exp\left[ 
Y_{\alpha} -\frac{p-1}{2} 
Tr\ln\underline{\underline{\Lambda}} \right] \langle\langle
\Theta(\tilde{r}_{\alpha\beta},\tilde{r}_{\alpha\alpha},m^{\alpha}_{1})\rangle\rangle_{\xi}
\end{split}
\label{med1}
\end{equation}
where $Y_{\alpha} = i\sum_{\alpha}{\tilde{r}}_{\alpha\alpha}q_{\alpha\alpha}+i\sum_{\alpha\neq\beta}
\tilde{r}_{\alpha\beta}q_{\alpha\beta} -\frac{\beta JN}{2}\sum_{\alpha}(m^{\alpha}_{1})^{2} $, 
$\Theta(\tilde{r}_{\alpha\beta},\tilde{r}_{\alpha\alpha},m^{\alpha}_{1})
=\mbox{Tr}_{s^{\alpha}} \exp{A^{\alpha}_{i}}$
with 
\begin{equation}
\begin{split}
 & A^{\alpha_{0}}_{i}= 
-i\sum_{\alpha\neq\beta}{\tilde{r}}_{\alpha\beta}(\frac{1}{N}
\sum_{i}S_{i}^{\alpha} S_{i}^{\beta}) +\beta J
\sum_{\alpha} (\sum_{i}\xi_{i}^{1}S_{i}^{\alpha})
m_{1}^{\alpha} 
 \\ &-\sum_{\alpha}(\frac{\beta J p}{2N} -\beta D 
 +\frac{i}{N}
 {\tilde{r}}_{\alpha\alpha})
 \sum_{i}
(S_{i}^{\alpha})^{2}
+ \frac{\beta^2 \Delta^2}{2} (\sum_{\alpha=1}^n S^{\alpha})^2 
\end{split}
\end{equation}
 and trace of the matrix $\underline{\underline{\Lambda}}$ obtained in terms of its eigenvalues.

The free energy is found introducing Eq. (\ref{med1}) in Eq. (\ref{replica}) which
is evaluated at the saddle point.
Thus,
\begin{equation} 
-i{\tilde{r}}_{\alpha\alpha}=\frac{\beta^{2}J^{2}}{2}\langle(m^{\alpha}_{1})^{2}\rangle=\frac{\beta^{2}J^{2}}{2}p\ r_{\alpha\alpha}  
\label{e27}
\end{equation}
and
\begin{equation} 
-i{\tilde{r}}_{\alpha\beta}=\frac{\beta^{2}J^{2}}{2}\langle (m^{\alpha}_{1}m^{\beta}_{1})\rangle=
\frac{\beta^{2}J^{2}}{2}p\ r_{\alpha\beta};~~~\alpha \neq \beta. 
\label{e27d}
\end{equation}

\section{The 1S-RSB procedure}\label{1s-rsb}

In the 1S-RSB procedure \cite{Parisi},  the replica 
matrix $\{Q\}$ and the matrix $\{r\}$ are parametrized as:
\begin{equation}
 \begin{array}{cc} 
 X_{\alpha \beta}=\left\{
\begin{aligned}
 \bar{X} \ \ \  &\mbox{if}\ \alpha=\beta
\\ X_1 \ \ \   &\mbox{if}\  I(\alpha/x)=I(\beta/x)
\\ X_0  \ \ \  & \mbox{if}\ I(\alpha/x)\neq I(\beta/x)
\end{aligned}\right.
\end{array}
\label{1srsb}
\end{equation}
where $X=q$ or $r$ and order parameters
$m^{\alpha}_{1}$ are invariant under permutation of replicas: $m^{\alpha}_{1}= m$, where 
$\alpha=1,\cdots n$.
 Parametrization (\ref{1srsb}) is used 
in Eq. (\ref{med1}) to obtain the 1S-RSB free energy as 
\begin{equation} 
\begin{split}
 &\beta f= B_a R_0 +\frac{\beta J m^2}{2} -\frac{1}{2}
 \frac{\beta J a q_0}{Q_0} +\frac{a}{2}\ln [Q_1]
\\& +\frac{a}{2x}\ln \frac{Q_0}{Q_1}
 -\lim_{n\rightarrow 0}\frac{1}{ n}\ln\langle\langle 
\Theta(\{r\},m,\xi) \rangle\rangle_{\xi},
\end{split}
\end{equation} 
where 
\begin{equation}
\begin{split}
 \Theta(\{r\},m,\xi)=\mbox{Tr}_{s^{\alpha}}  \exp\left[ O_{\alpha} + N_{\alpha} \right]
\end{split}
\end{equation}
with 
\begin{equation}
 O_{\alpha} = B_a r_0(\sum_{\alpha=1}^n S^{\alpha})^2 + B_{d} (\sum_{\alpha=1}^n S^{\alpha})^2   
+\beta J \sum_{\alpha=1}^{n} \xi m S^{\alpha}
\end{equation}
\begin{equation}
 N_{\alpha} =  B_a \bar{R} \sum_{\alpha=1}^n(S^{\alpha})^2  + B_a R_1\sum_{l=1}^{n/x}(\sum_{\alpha=(l-1)x+1}^{lx}S^\alpha)^2
\end{equation}
and $B_a= \frac{\beta^2J^2a}{2}$, $B_{d} = \frac{\beta^2 \Delta^2}{2}$,  $\bar{R}= (\bar{r}-r_1-1/\beta J) -\beta D$, $R_1=(r_1-r_0)$ and $a=p/N$. 
The quadratic forms into the function 
$\Theta(\{r\},m,\xi)$ can be 
linearized by Hubbard-Stratonovich transformations 
where new auxiliary fields are introduced in the problem. Therefore, one has
\begin{equation} 
\begin{split}
&\Theta(\{r\},m,\xi)=
\int  Dz
\left[
\int  Dv
\left(\int  Dw 
\mbox{Tr}_{S} e^{\Xi}\right)^x 
\right]^{n/x} ,
\end{split} 
\label{e28}
\end{equation}
with $Dy= \frac{dy e^{-\frac{y^{2}}{2}}}{\sqrt{2\pi}}$ ($y=z,\ v,\ w$) and

\begin{equation} 
\begin{split}
\Xi&=\beta J[ \sqrt{ar_0 + (\Delta/J)^2}z+\sqrt{a(r_1-r_0)}v
\\&+\sqrt{a(\bar{r}-r_1-1/\beta J) - 2D/J}w +\xi m]S.
\end{split}
\label{e30}
\end{equation}

 The 1S-RSB free energy can then be expressed by 
Eq. (\ref{eq1}), in which the order parameters $m,\ q_0,\ q_1$, and $\bar{q}$ and the replica block size parameter $x$ are given by the saddle point equations:
\begin{equation}
 m=\int Dz\langle\langle  \xi\frac{\int Dv K(z,v|\xi)^{x-1} 2
e^{\gamma} \sinh H }{\int Dv K(z,v|\xi)^x  }\rangle\rangle_{\xi}
\end{equation}
\begin{equation}
 q_0=\int Dz \langle\langle \left(\frac{\int Dv K(z,v|\xi)^{x-1}2 e^{\gamma}
 \sinh H}{\int Dv K(z,v|\xi)^x  }\right)^2\rangle\rangle_{\xi}
\end{equation}
\begin{equation}
 q_1=\int Dz\langle\langle  \frac{\int Dv K(z,v|\xi)^{x-2} \left[ 2 e^{\gamma} \sinh H \right]^2}{\int Dv K(z,v|\xi)^x  }\rangle\rangle_{\xi}
\end{equation}
\begin{equation}
\bar{q}=\int Dz\langle\langle \frac{\int Dv K(z,v|\xi)^{x-1}2 e^{\gamma} \cosh H
}{\int Dv K(z,v|\xi)^x} \rangle\rangle_{\xi}
\end{equation}
and
\begin{equation}
\begin{split}
\frac{1}{x}\int Dz\langle\langle \ln \int Dv K(z,v|\xi)^{x} \rangle\rangle_{\xi}
\\
-\frac{1}{x}\int Dz\langle\langle  \frac{\int Dv K(z,v|\xi)^{x}\ln K(z,v)^{x} }{\int Dv K(z,v|\xi)^x}\rangle\rangle_{\xi}
\\
+\frac{\beta J a}{2}\left(\frac{q_0}{Q_0}-\frac{q_1}{Q_1}\right)
-\frac{a}{2x}\ln\frac{Q_0}{1-\beta J Q_1}
=0 
\end{split}
\end{equation}
where $K(z,v|\xi)$ and $H=H(z,v|\xi)$ are defined in Eqs. (\ref{Kzv}) and (\ref{eq20}), respectively.

\hspace{3cm}


\begin{thebibliography}{00}

\bibitem{Dotsenko} V. Dotsenko, 2001 {\it Introduction to the Replica Theory of Disordered Statistical Systems} (Cambridge: Cambridge University Press)
\bibitem{Nishimori} H. Nishimori, 2001 {\it Statistical Physics of Spin Glasses and Information Processing} (Oxford: Oxford University Press)
\bibitem{Young} A. P. Young (ed), 1998 {\it Spin Glasses and Random Fields} (Singapore: World Scientific)
\bibitem{Belanger1} D. P. Belanger, {\it  Experiments on the random field Ising model }, 1998 {\it Spin Glasses and Random Fields} ed. A. P. Young (Singapore: World Scientific)
\bibitem{Belanger2} D. P. Belanger, Wm. E. Murray Jr., F. C. Montenegro, A. R. King, V. Jaccarino and R. W. Erwin, 1991 Phys.
Rev. B {\bf 44} 2161; F. C. Montenegro, A. R. King, V. Jaccarino, S-J. Han and D. P. Belanger, 1991 Phys. Rev. B {\bf 44} 2155
\bibitem{JMMM2013} S. G. Magalhaes, F. M. Zimmer, B. Coqblin, J. Magnetism Magnetic Materials {\bf 226-230}, 148 (2013).
\bibitem{Nattermann1} T. Nattermann and J. Villain, 1988 Phase Transit. {\bf 11} 5
\bibitem{Nattermann2} T. Nattermann, {\it Theory of the random field Ising model}, 1998 {\it Spin Glasses and Random Fields} ed. A. P. Young (Singapore: World Scientific)
\bibitem{Dotsenko2} V. Dotsenko, 2007 J. Stat. Mech. {\bf 2007} P09005
\bibitem{SNA} R. F. Soares, F. D. Nobre and J. R. L. de Almeida, 1994 Phys. Rev. B {\bf 50} 6151
\bibitem{NNCC} E. Nogueira, F. D. Nobre, F. A. da Costa and S. Coutinho, 1998 Phys. Rev. E {\bf 57} 5079; E. Nogueira, F. D. Nobre, F. A. da Costa, and S. Coutinho, 1999 Phys. Rev. E {\bf 60} 2429 (erratum)
\bibitem{ANC} J. M. de Ara\'ujo, F. D. Nobre and F. A. da Costa, 2000 Phys. Rev. E {\bf 61} 2232
\bibitem{CN} N. Crokidakis and F. D. Nobre, 2008 Phys. Rev. E {\bf 77} 041124
\bibitem{MMN} S. G. Magalh\~aes, C. V. Morais and F. D. Nobre, 2011 J. Stat. Mech. {\bf 2011} P07014
\bibitem{SK} D. Sherrington and S. Kirkpatrick, Phys. Rev. Lett. {\bf 35} 1792 (1975).
\bibitem{ghatak} S. K. Ghatak and D. Sherrington, J. Phys. C {\bf 10}, 3149 (1977). 
\bibitem{amit} D. J. Amit, {\it Modelling Brain Function. The World of Attractor Neural Networks} (Cambridge University Press, Cambridge, England, 1989).
\bibitem{amit2} D. J. Amit, H. Gutfreund, and H. Sompolinsky, Ann. Phys. {\bf 173} 30 (1987).
\bibitem{0} B. Donnio et al, Softmatter {\bf 6}, 965 (2010).
\bibitem{1} F. Jona and G. Shirane, 1962 {\it Ferroeletric Crystals} (New York: Pergamon)
\bibitem{2} {B. \ifmmode \check{Z}\else \v{Z}\fi{}ek\ifmmode \check{s}\else \v{s}\fi{}}, G. C. Shukla, and R. Blinc, 1971 {\it Phys. Rev. B} {\bf 3} 2306.
\bibitem{3} P. E. Cladis, R. K. Bogardus, W. B. Daniels and G. N. Taylor, 1977 {\it Phys. Rev. Lett}
{\bf 39} 720
\bibitem{4} P. E. Cladis, D. Guillon, F. R. Bouchet and P. L. Finn, 1981 {\it Phys. Rev. A} {\bf 23}
2594
\bibitem{5} S. Rastogi, G. W. H. Hohne and A. Keller, 1999 {\it Macromolecules} {\bf 32} 8897;\\
N. J. L. van Ruth, S. Rastogi, 2004 {\it Macromolecules} {\bf 37} 8191
\bibitem{6} O. Portmann, A. Vaterlaus, D. Pescia, 2001 {\it Nature} {\bf 422} 701
\bibitem{7} N. Avraham, B. Khaykovich, Y. Myasoedev, M. Rapoport, H. Shtrikman, D. E.
Feldman, T. Tamegai, P. H. Kes, M. Li, M. Konczykowski, K. van der Beek and E. Zeldov, Nature
{\bf 411} 451 (2001).
\bibitem{8} A. Scholl et al., Science {\bf 329}, 303 (2010).
\bibitem{Debenedetti} M. R. Feeney, P. G. Debenedetti, F. H Stillinger, J. Chem. Phys. {\bf 119}, 4582 (2003).
\bibitem{9} N. Schupper, N. M. Shnerb, Phys. Rev. Lett. {\bf 93} 037202 (2004);\\
N. Schupper, N. M. Shnerb, Phys. Rev. E {\bf 72} 046107 (2005).
\bibitem{10} M. Paoluzzi, L. Leuzzi and A. Crisanti, Phys. Rev. Lett. {\bf 104} 120602 (2010);\\
L. Leuzzi, M. Paoluzzi and A. Crisanti, Phys. Rev. B. {\bf 83} 014107 (2011).
\bibitem{11} A. Crisanti and L. Leuzzi, Phys. Rev. Lett. {\bf 95} 087201 (2005).
\bibitem{12} F. A. da Costa, Phys. Rev. B {\bf 82} 052402 (2010).
\bibitem{13} F. M. Zimmer, C. F. Silva, C. V. Morais, S. G. Magalhaes, J. Stat. Mech. {\bf 2011},
05026 (2011).
\bibitem{14} S. G. Magalh\~aes, C. V. Morais and F. M. Zimmer, Phys. Rev. B {\bf 81} 014207
(2010).
\bibitem{15} S. G. Magalh\~aes, C. V. Morais and F. M. Zimmer, Phys. Rev. B {\bf 77} 134422
(2008).
\bibitem{16} C. V. Morais, F. M. Zimmer and S. G. Magalh\~aes, Phys. Lett. A {\bf 375} 689
(2011).
\bibitem{17} C. K. Thomas, H. G. Katzgraber, Phys. Rev. E {\bf 84}, 040101(R) (2011).
\bibitem{16b} C. V. Morais, M. J. Lazo, F. M. Zimmer and S. G. Magalh\~aes, Physica A {\bf 392} 1770 (2013).
\bibitem{16a} C. V. Morais, M. J. Lazo, F. M. Zimmer and S. G. Magalh\~aes, Phys. Rev. E {\bf
85} 031133 (2012).
\bibitem{fradkin} E. Fradkin, B. A. Huberman, and S. H. Shenker, Phys. Rev. B {\bf 18} 4789 (1978).
\bibitem{binder} K. Binder and A. P. Young, Rev. Mod. Phys. {\bf 58} 801 (1986).
\bibitem{provost} J. P. Provost and G. Vallee, Phys. Rev. Lett. {\bf 50} 598 (1983).
\bibitem{Nobre1} R. F. Soares, F. D. Nobre, and J. R. L. de Almeida, Phys. Rev. B {\bf 50}, 6151 (1994).
\bibitem{Parisi} G. Parisi, J. Phys. {\bf 13}, 1101 (1980). 
\bibitem{Castillo} I. P. Castillo, D. Sherrington, Phys. Rev. B {\bf 72}, 104427 (2005).
\bibitem{bolle} D. Bolle, H. Rieger and G. M. Shim, J. Phys. A: Math. Gen. {\bf 27}, 3411 (1994).
\bibitem{qiang} Y. Q. Ma, Y. M. Zhang, and C. D. Gong, Phys. Rev. B {\bf 46}, 11591 (1992). 
\bibitem{Coqblin1} S. G. Magalhaes, F. M. Zimmer, P. R. Krebs, and B. Coqblin, Phys. Rev. B {\bf 74}, 014427 (2006).
%
\bibitem{Coqblin2} S. G. Magalhaes, F. M. Zimmer, B. Coqblin, Phys. Rev. B {\bf 81}, 094424 (2010).

\end{thebibliography}
\end{document}